\documentclass[12pt]{article}
\usepackage[hmargin=2.5cm,vmargin=2.5cm]{geometry}
 \usepackage{graphicx}
 \usepackage{amsmath}
 \usepackage{amssymb}
  \usepackage{enumerate}
\usepackage{cite}
\usepackage{mathabx}
\usepackage{url}
\newcommand{\bey}{\begin{eqnarray}}
\newcommand{\eey}{\end{eqnarray}}
\usepackage{etoolbox}
\usepackage{tabularx}
\usepackage{lipsum}
\usepackage{MnSymbol,wasysym}
 
\newcommand{\csch}{\mathrm{csch} \,}

\newcommand{\R}{\mathbb{R}}

\begin{document}

 \title{Including gravity in equilibrium thermodynamics }

 \author {  Eirini Sourtzinou\footnote{sourginou@upatras.gr}   \; and Charis Anastopoulos\footnote{anastop@upatras.gr}\\
Laboratory of Universe Sciences, Department of Physics, \\University of Patras, 26504 Patras, Greece.}

\maketitle

\begin{abstract}
This paper is part of a bottom-up approach to gravitational thermodynamics that is guided by the axiomatic frameworks of equilibrium thermodynamics. We identify a novel form of the microcanonical distribution for systems in background gravitational fields that respects the kinetic theory  and the thermodynamic symmetries. Thermodynamic  consistency dictates the treatment of the gravitational field as a thermodynamic variable. We introduce the thermodynamic conjugate to the gravitational field, the gravitational pull, an additive variable  that  is a structural element of our microcanonical distribution. We demonstrate the validity of our results to inhomogenous background fields, a class of self-gravitating systems, relativistic gases in Rindler spacetime, and quantum gases.

\end{abstract}

\section{Introduction}
The interplay between gravity and thermodynamics is a multi-faceted problem in the foundations of physics. Its aspects include black hole thermodynamics \cite{Bek1, Bek2, BCH, Hawk1, Wald}, the possibility of gravitational entropy with a cosmological imprint \cite{Penrose, Wald2, Wallace, Kron}, and the challenge posed by 
 the non-extensivity of self-gravitating systems to the traditional accounts of statistical mechanics \cite{Pad, Katz, Chavanis, AnSav14}.  There is no unifying framework for those different domains, even when restricting to the case of equilibrium.

This work is part of a bottom-up approach to the topic that focuses  on axiomatic formulations of equilibrium thermodynamics \cite{Caratheo, Landsberg, Giles, Callen, Lieb}. The idea is to use the feedback from models in gravitational thermodynamics in order to construct a broader axiomatic framework  to include gravitational effects. This will likely involve abandoning some principles---such as the extensivity of entropy---of the current theory, and the introduction of additional structures, such as the concept of the gravitational pull that we introduce in this paper.
 
Our current models for gravitational thermodynamics  can be hierarchized in ascending order of difficulty, as follows.
 
\begin{enumerate}
\item thermodynamic systems in a background gravitational field;
\item self-gravitating systems in a background spacetime (e.g., via Newtonian gravity);
\item self-gravitating systems in full General Relativity;
\item self-gravitating systems that incorporate quantum effects (e.g., Hawking radiation).
 
\end{enumerate}
Each level in the hierarchy introduces a host of novel problems, both conceptual and technical. Our long-term aim is to identify a single axiomatic framework that will be valid at all levels. To find this framework, we must start at the bottom level and  climb up.    For other works that emphasize the axiomatic approach to thermodynamics for   gravitational systems, see Refs. \cite{Martin1, Martin2, AnSav12, Kotop1}.

This paper works at the  first and second levels of the hierarchy. We argue that a consistent thermodynamic description requires the treatment of a background gravitational field as a thermodynamic variable, the same way that an external magnetic field is treated as a thermodynamic variable in magnetic systems. This implies that the thermodynamic conjugate  of the field, the {\em gravitational pull $Q$}, must be a variable of the fundamental space of the system.   In the simplest systems, the gravitational pull is an additive version of the center of mass. Hence, the presence of gravity requires the introduction of new thermodynamic variables.

An external field breaks entropy extensivity even in the absence of self-gravity. Therefore, we have to introduce   variables other than the volume to describe the region in which the system is enclosed. This   results in different pressures exerted at different directions. In Refs. \cite{SouAn23, Souphd}, it was shown that such effects are important on black hole backgrounds, as they lead to ``buoyant" forces \cite{UW} near the horizon. 

The correct identification of the internal energy is a major issue for self-gravitating systems \cite{AnSav14}. Here we demonstrate that the internal energy differs from the total energy of a system. 
In particular,   the internal energy  of ideal gases  should be identified with the kinetic energy of the molecules, in accordance with kinetic theory.

Our methodology in this paper involves a mixture of purely thermodynamic arguments and simple, analytically tractable statistical mechanics models.  Our results include the following. 

\begin{itemize}
\item We provide a full thermodynamic analysis of the paradigmatic system of a box of gas in an external homogeneous gravitational field, by including the gravitational pull and the pressure inhomogeneity in the thermodynamic description (Sec. 2). This leads to a novel microcanonical distribution for the system, and a reinterpretation of the canonical distribution \cite{LDP} (Sec. 3).

\item Our thermodynamic analysis fully applies  to self-gravitating systems, as we demonstrate by analyzing the one-dimensional analogue of the isothermal sphere \cite{BiTr, Chandr}. This model enables a straightforward comparison of the gravitational pull to  polarization and magnetization in condensed matter. (Sec. 4).

\item The generalization to quantum gases is straightforward. We find a genuine phase transition for fermions in a background field, the phases corresponding to 
 whether 
 the gas reaches the top of the container or not.

\item Our analysis also applies to inhomogeneous gravitational fields (Sec. 5),  and to relativistic systems (Sec. 7). The latter result follows from the analysis of ideal gases in Rindler spacetime \cite{SouAn23, LMar, Pad2, Pad3}. 
\end{itemize}


\section{Thermodynamics of a non-relativistic gas in a gravitational field}

We start our analysis with the simplest thermodynamic system that is affected by gravity: a classical gas  in a static box within a constant gravitational field. 

At the microscopic level, this system is described by the Hamiltonian of 
 $N$ particles of masses $m_i$ in a constant gravitational field ${\bf g}$ 
\bey
H = \sum_{i=1}^N \frac{{\bf p}_i^2}{2m_i} + \sum_{i=1}^N \sum_{j < i}  V({\bf x}_i - {\bf x}_j) +\sum_{i=1}^N m_i {\bf g}\cdot ({\bf x}_i - {\bf a}), \label{Hammm}
\eey
where $V$ is the  potential for particle interaction, and ${\bf a}$ an arbitrary constant that reflects our freedom to choose the zero of the gravitational potential.

We define 
\bey
{\bf Q} = \sum_{i=1}^N m_i ({\bf x}_i - {\bf a}), \label{Q}
\eey
the total mass $M = \sum_{i=1}^N m_i$ and the center-of-mass velocity ${\bf V}^c = \sum_{i=1}^N {\bf p}_i/m_i$. The quantity ${\bf Q}$ is canonically conjugate to ${\bf V}$, as their Poisson bracket is $\{Q_i, V_j^c\} = \delta_{ij}$. 
 The Hamiltonian separates as
\bey
H = H_0 + \frac{1}{2} M {\bf V}^2 + {\bf g} \cdot {\bf Q}, \label{HH0}
\eey
where $H_0$ is the Hamiltonian  at the system's rest frame. Equivalently, we can use the conjugate pair ${\bf X} = {\bf Q}/M$ and ${\bf P} = M {\bf V}$, which corresponds to the center of mass coordinate and its conjugate momentum. 


\subsection{The fundamental representation}

In axiomatic formulations of equilibrium thermodynamics, the starting point is the identification of the fundamental space $\Lambda$. This consists of the variables that  specify the spatial boundary of the system, and by additive conserved quantities \cite{Callen, Callen2}. The latter include particle numbers $N_a$ for the different particle species, and the internal energy $U$. In general, a thermodynamic system may involve multiple components.  For example, it may consist of  two boxes in contact through semi-permeable  or moveable walls. Thermodynamic properties are defined in terms of the entropy functional $S: \Lambda \rightarrow \R^+$. 

In absence of a external fields, and for sufficiently large particle numbers, the geometric properties of the boundary are irrelevant to the thermodynamic description: only the total volume $V$ that is enclosed by the boundary contributes to the entropy. Hence, for  a gas with a single particle species, the fundamental space $\Lambda$ consists of three variables $U, N$, and $V$, and the entropy function satisfies the extensivity property
\bey
S(tN, tU, tV) = t S(N, U, V), \hspace{0.5cm} \mbox{for all} \; t > 0. \nonumber
\eey

Suppose that the  box of gas is placed within an external gravitational field ${\bf g} = (0, 0, g)$, where it remains static by the action (${\bf V}^c = 0$) of an external force. We identify the fundamental thermodynamic space through the following considerations.

\begin{enumerate}
\item The external field breaks space isotropy, so the volume $V$ is not the only spatial variable that  describes the system. Consider a rectangular box with sides $L_1, L_2$, and $L_3 = L$. Due to translation symmetry in the 1-2 plane, thermodynamic quantities depend only on the product $A = L_1L_2$, and not on $L_1$ and $L_2$ separately. But there is not translation symmetry in the 3-direction, so $L$ is an independent thermodynamic quantity.
    
\item When considering a single box of gas, we can always choose the coordinate origin at the geometric center of the box.
 However, in a system of two boxes, the relative coordinate of their geometric center matters, because  the higher placed box has more potential energy, which can be used to generate work. For this reason, we must include the 3-coordinate $\ell_0$ of the geometric center of the box into the thermodynamic description. Rather than $L$ and $\ell_0$, we can use the coordinates $\ell_b = \ell_0 - L/2$ for the bottom of the box and $\ell_t =  \ell_0 + L/2$ for the top of the box. This enables us to define the pressures $P_t$ at the top of box, $P_b$ at the bottom, and the horizontal pressure $P_h$, through work terms $dW_t = - P_t A d\ell_t$, $dW_b=  P_b A d\ell_b$, and $dW_h =   -  P_h L dA$, respectively. 
 
 \item A homogeneous gravitational field couples to the center of mass of the gas, and the associated work term is $dW = - g dQ$, where we wrote ${\bf Q} = (0, 0, Q)$. Hence, $Q$ is the thermodynamic conjugate to the gravitational acceleration $g$. It is convenient to take the arbitrary vector ${\bf a}$ in Eq. (\ref{Q}) to coincide with the point vector of the geometric center of mass, so that  $Q = \sum_{i=1}^N m_i (x_{3i} - \ell_0) $. Then, $Q$ is an additive quantity (scaling with the number $N$ of particles) that vanishes in absence of the gravitational field.  $Q$ measures the difference of the  center of mass from the geometric center of the box, hence, it is a measure of the gravity-induced inhomogeneity of the gas. For this reason, we will refer to $Q$ as the {\em gravitational pull} of the system.
     
    The gravitational pull is the direct analogue of the polarization and the magnetization, for dielectric and magnetic systems, respectively. For a dielectric system in an electric field ${\bf E}$, polarization is defined by ${\bf P} =  \sum_{i} q_i {\bf x}_i$, where $q_i$ are the particle charges. The polarization is a variable on the fundamental space, as it corresponds to a work term   $dW = - {\bf E}\cdot d {\bf P}$.

 \end{enumerate}

We conclude that the fundamental thermodynamic space $\Lambda$ of a box of gas in the gravitational field consists of the variables $U, N, A, L, Q, \ell_0$, defined earlier. For a single box of gas, the entropy does not depend on $\ell_0$, but we must keep track of this variable when dealing with systems of two or more boxes at different heights within the gravitational field. 

The only remaining issue is to identify the internal energy $U$. There are two candidates: the energy associated to the Hamiltonian $H_0$ and the energy associated to the Hamiltonian $H$ in Eq. (\ref{HH0}). The correct choice is the first one. The reasons are the following.

\begin{enumerate}

\item The difference between $H$ and $H_0$ is a term proportional to the center of mass of the system. Center of mass degrees of freedom are not included in the internal energy, either in equilibrium or in non-equilibrium thermodynamics. \cite{LL, GrMa}.

\item When applying Boltzmann's kinetic theory to an ideal gas in a gravitational field, the temperature depends only on the average kinetic energy of the molecules and not on the average of their total energy \cite{CoLa}, suggesting that the kinetic energy is to be identified with the internal energy.

\item For non-relativistic systems, we expect that two identical boxes of gas held at different heights have the same temperature. However, the total energy of the higher box is larger. If the internal energy were identified with the total energy, the higher box would be hotter. 
\end{enumerate}

With this identification, the first law of thermodynamics becomes
\bey
dU = TdS - P_h L dA + P_b A d\ell_b - P_t A d\ell_t + \mu d N - g d(Q + M \ell_0) \nonumber \\ = 
TdS - P_h L dA - P_v A dL +  \mu d N - g dQ. \label{rep1}
\eey
The requirement that the entropy does not depend on $\ell_0$ implies the following relations between the ``vertical" pressure $P_v$  and  the pressures $P_t$ and $P_b$,
\bey
P_t = P_v - \frac{Mg}{2A}, \hspace{1cm}P_b = P_v + \frac{Mg}{2A}.
\eey
The pressures $P_v$ and $P_h$ are, in general,  different.  The horizontal pressure is obtained from the average force exerted on the walls that are parallel to the acceleration vector. This force is not distributed equally on the wall, as the gas is inhomogeneously distributed. This does not mean necessarily  that the local pressure, as defined by the stress-energy tensor, is anisotropic. To avoid confusion, we will say that the system is characterized by  asymmetric pressure. 

To  monitor the change of pressure with height, we should have work terms that correspond to pressures at different heights. This is not possible with rigid walls, we should have to enlarge the thermodynamic space to include general surface deformations.

We need to distinguish the degenerate limit, at which the entropy does not depend on the length $L$. This corresponds, for example, to the case in which $\ell_t \rightarrow \infty$, i.e., to a semi-infinite box. In this case, we choose the vector ${\bf a}$ in ${\bf Q}$ to coincide with $(0, 0, \ell_b)$. The invariance of the entropy under $\ell_b$ implies that $P_b = Mg/A$, while $P_t = 0$.

\subsection{Temperature}
The gas is not distributed homogeneously within the box in presence of a gravitational field. Nonetheless, the temperature remains constant within the body, at least in the non-relativistic limit.  This can be demonstrated through the maximum entropy principle.

Assume that the system is in local equilibrium with entropy density $s(x)$ that is a local function of the number density $n(x)$ and the energy density $u(x)$, $s(x) = s[u(x), n(x)]$. The total entropy is $S = \int  d  x \, s[u(x), n(x)]$. The forms of $u(x)$, $n(x)$ are constrained by the maximum entropy principle: the entropy $S$ is maximum for constant internal energy $U = \int dx \, u(x)$, particle number $N = \int dx \, n(x)$, and gravitational pull $Q(x) = m 
\int dx \, x  n(x)$. Hence, we maximize the quantity $S + \beta U + \gamma N + \eta  Q$ with respect to variations in $u(x)$ and $n(x)$, $\beta, \gamma,$ and $\eta$ are Lagrange multipliers..
We obtain
\bey
\frac{\partial s}{\partial u} = \beta, \hspace{1cm}
\frac{\partial s}{\partial n} = \gamma + m \eta x. \nonumber
\eey
When assuming local equilibrium, ${\partial s}/{\partial u}$ is identified with the local temperature $T(x)$, and ${\partial s}/{\partial n}$ with the ratio $\mu(x)/T(x)$, where $\mu(x)$ is the local chemical potential. Then, the first equation above implies that the local temperature  
  $T(x) = \beta^{-1}$ is   constant. The second equation  yields $\mu(x)/T(x) = \gamma + m \eta x$, or equivalently $\mu(x) = \gamma/\beta + m (\eta/\beta) x$. This is the  {\em Gibbs formula} for the chemical potential \cite{Gibbs}, provided that we identify $\eta/\beta$ with the gravitational acceleration. 
  

\subsection{Other thermodynamic potentials}

In the fundamental space, we can employ either  the entropy
representation, where the entropy $S$ is  function of the variables $U, N, A, L,$ and $Q$; or with the internal energy representation, where 
 $U$ is a function of $N$, $S$, $A$, $L$, and $Q$. In the latter representation, $g = - \left(\partial U/\partial Q\right)_{N, S, A, L}$.

The Legendre transform of $U$ with respect to $Q$ is the {\em total energy} $E = U + g \, Q$, which is a state function of  $N$, $S$, $A$, $L$, and $g$. The first law of thermodynamics in the total energy representation reads
\bey
dE = TdS - P_h L dA - P_v A d L + \mu d N + Q dg. \label{rep2}
\eey
Eq. (\ref{rep2}) implies that entropy $S$ can also be expressed as a state function of the total energy $E$ and the gravitational acceleration $g$. This is {\em not} the fundamental representation, because it involves the intensive variable $g$, rather than its extensive conjugate $Q$.

The Legendre transform of the internal entropy $U$ with respect to the entropy $S$ yields the {\em Helmholtz free energy} $F = U - TS$, as a function of $N$, $T$, $A$, $L$, and $Q$. The Legendre transform of the total energy $E$ with respect to $S$, yields the {\em Gibbs free energy} \ $G  = U - TS + g\, Q$ which is a state function of  $N$, $T$, $A$, $L$, and $g$. 

We also define two versions of the Landau potential. The standard Landau potential $\Phi = F - \mu N$ is defined as a Legendre transform of the Helmholtz free energy with respect to $N$, and it is a function of $\mu$, $T$, $A$, $L$, and $Q$. The transformed Landau potential $\tilde{\Phi} = \tilde{F} - \mu N$ is the Legendre transform of $G$ with respect to $N$, and it is a function of $\mu$, $T$, $A$, $L$, and $g$.

\subsection{Thermodynamic quantities}
The inclusion of the gravitational field into the thermodynamic description and the asymmetry of pressure enable the definition of new, operationally accessible thermodynamic quantities. These quantities are {\em responses}, they record how an extensive variable responds to a change in an intensive variable.

First, we define the {\em gravitational susceptibility} at constant temperature (an analogue of the magnetic susceptibility) as
\bey
\chi_T = -\frac{1}{N } (\partial Q/\partial g)_{N, T, L, A}.
\eey
We also define different versions of compressibility at constant temperature, depending on the pressure that is being varied and the spatial direction---horizontal (h) or vertical (v)---whose response we monitor: 
\bey
\kappa_{T}^{vv} = -\frac{1}{L}(\partial L/\partial P_v)_{N, g, T}, \hspace{1cm}
\kappa_{T}^{vh} = -\frac{1}{A}(\partial A/\partial P_v)_{N, g, T}, \nonumber \\
\kappa_{T}^{hv} = -\frac{1}{L}(\partial L/\partial P_h)_{N, g, T}, \hspace{1cm}
\kappa_{T}^{hh} = -\frac{1}{A}(\partial A/\partial P_h)_{N, g, T}.
\eey
Finally, we   define the gravity-induced asymmetry as $\Delta = (P_v - P_h)/P_h$, and the {\em asymmetry index} as the rate of change of $\Delta$ with respect to the field $g$,
\bey
\zeta_T = \left(\frac{\partial \Delta}{\partial g}\right)_{N, T, L, A}.
\eey

\section{Statistical mechanics of a non-relativistic gas in a gravitational field}
\subsection{The microcanonical distribution}
In this section, we will analyze the statistical mechanics of gases in a gravitational field. The first step is to use the microcanonical distribution, in order to construct the fundamental representation from the microscopic dynamics. However, this step crucially depends on the correct identification of the fundamental space. Past work on the topic did not identify the gravitational pull as a thermodynamic variable, and this led to a different expression for the microcanonical distribution from the one that we employ here. 

In particular, Refs. \cite{RWV, RGWV} employ  a microcanonical distribution in which they identify the total energy with the internal energy
\bey
\rho_{E, g} (x, p) = \tilde{\Gamma}(E, N, g)^{-1} \delta(H_0 + m g \sum_{i=1}^n x_{3i} - E) \delta({\bf P}), \label{diseg}
\eey
where 
\bey
\tilde{\Gamma}(E, N, g) = \int \frac{d^{3N}x d^{3N}p}{(2\pi \hbar)^{3N}N!}   \delta(H_0 + m g \sum_{i=1}^n x_{3i} - E) \delta({\bf P}) \label{tildeg}
\eey
is the volume of the energy surface. 
One may attempt to define the entropy  $\tilde{S}(E, N, g) =  \log \tilde{\Gamma}(E, N, g)$. However, the microcanonical distribution is supposed to be defined on the fundamental space, while $\tilde{S}$ is not. 

For an ideal gas, Eq. (\ref{tildeg}) yields
\bey
\tilde{\Gamma}(E, N, g) = \frac{1}{L^N}\int  d^N x_3  \Gamma^{(0)}\left(E - mg\sum_{i=1}^N x_{3i}, N\right), \label{geng}
\eey
where $\Gamma^{(0)}(U,N)$ is the volume of the energy surface in absence of the gravitational field. In this case,  $E$ coincides with  $U$. Eq. (\ref{geng})  is problematic because it gives a temperature $T^{-1} = \partial S/\partial E$ that depends on the gravitational field. Furthermore, the gravitational contribution to the entropy does not factorize, despite the fact that the degrees of freedom coupled to the gravitational field  factor out in the dynamics.

Following up on our previous analysis,  we describe this system by a microcanonical distribution defined with reference to the fundamental thermodynamic space $\Lambda$. To this end, we demand constant values of the internal energy $U$ and the gravitational pull $Q$, 
\bey
\rho_{U, Q} (x, p) = \Gamma(U, N, Q)^{-1} \delta(H_0 - U) \delta(   \sum_{i=1}^n x_{3i} - Q/m) \delta({\bf P}), \label{microuq}
\eey
where 
\bey
\Gamma(U, N, Q) = \int \frac{d^{3N}x d^{3N}p}{(2\pi \hbar)^{3N}N!}   \delta(H_0  - U) \delta(   \sum_{i=1}^n x_{3i} - Q/m)  \delta({\bf P}).
\eey
The associated entropy  is  $S(U, N , Q) = \log \Gamma(U, N, Q)$. The distribution (\ref{microuq}) is not equivalent to (\ref{diseg}). The distribution associated to (\ref{microuq}) with reference to the total energy space is 
\bey
\rho_{E, \eta} (x, p) \sim \delta(H_0 - U) \exp[-\eta   \sum_{i=1}^n x_{3i} - Q/m)] \delta({\bf P}), \label{microuq2}
\eey
where $\eta$ is a constant, eventually to be identified with $g/T$. 

Regarding Eq. (\ref{microuq}), we note that
the microcanonical distribution usually follows from an ergodicity assumption that the only conserved quantity is the energy. However, in the present case, the center of mass coordinate completely factorizes from the remaining degrees of freedom, and it has vanishing Poisson bracket with the Hamiltonian $H_0$ (for ${\bf P} = 0$). Hence, it must appear as a separate independent variable describing the equilibrium state, in agreement with the thermodynamic analysis of Sec. 2.

Note that our arguments for the microcanonical distribution (\ref{microuq}) also  applies to self-gravitating systems. The microcanonical distribution for the latter ought to involve a delta function for the gravitational pull in addition to that for the internal energy.
 
For an ideal gas, we find that the volume function $\Gamma$ factorizes
\bey
\Gamma(U, N, Q) = \Gamma^{(0)}(U, N) \gamma(Q), \label{gunq}
\eey
where 
\bey
\gamma(Q) = \frac{1}{L^N} \int d^Ny  \delta( \sum_{i=1}^N y_{i} - Q/m).
\eey

We evaluate $\gamma(Q)$ as follows. 
\bey
\gamma(Q) &=& \frac{1}{2\pi L^N} \int d^Ny \int dk e^{ik(\sum_{i=1}^N y_{i} - Q/m)} = \frac{1}{2\pi L^N} \int_{-\infty}^{\infty}  dk e^{-ikQ/m} \left(\int_{-L/2}^{L/2}dy e^{iky}    \right)^N \nonumber 
\\
&=& \frac{1}{2\pi L^N} \int_{-\infty}^{\infty}  dk e^{-iNkq} \left( \frac{2 \sin(kL/2)}{k}\right)^N,
 \label{fq}
\eey
where  we took $\ell_0 = 0$. The quantity   $q = \resizebox{0.039\textwidth}{!}{$\frac{Q}{Nm}$}$ defines the gravitational pull per unit mass.   

\bigskip

\begin{figure}[tb] 
\includegraphics[height=6cm]{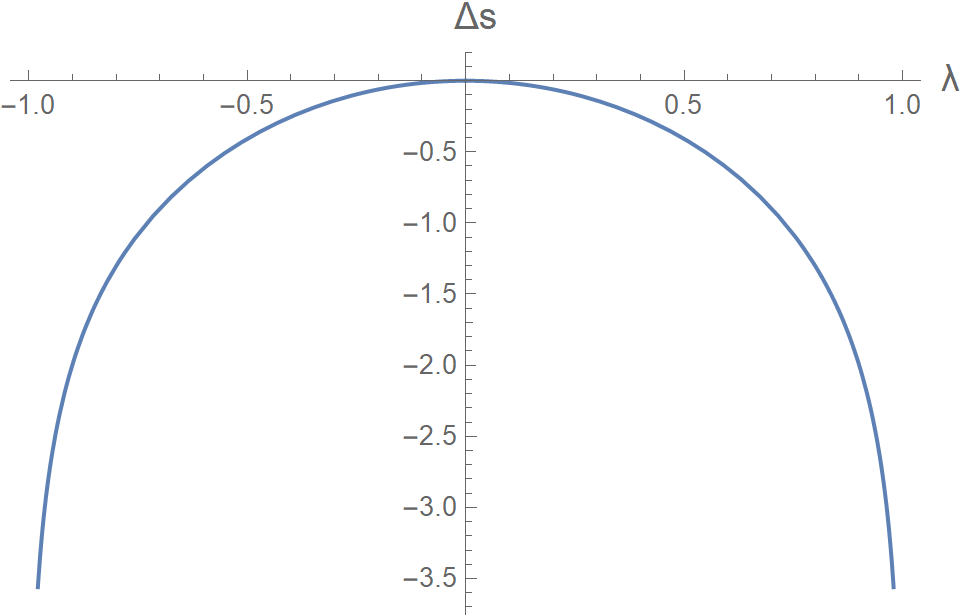} \caption{We plot the entropy per particle $\Delta s = \Delta S/N$ due to the gravitational field as a function of the dimensionless gravitational pull $\lambda = 2q/L$.}
\label{entropy1}
\end{figure}

We change the integration variable to $t = kL/2$. We obtain $\gamma(Q) = \resizebox{0.035\textwidth}{!}{$\frac{1}{\pi L}$}  \int_{-\infty}^{\infty} dt e^{-Nw(t)}$,
where
\bey
w(t) =  i \lambda t - \log\left(\frac{\sin t}{t}\right),
\eey
and $\lambda = 2q/L$. In the saddle-point approximation,  
\bey
\gamma(Q) = \sqrt{\frac{2}{\pi NL^2((\sinh b)^{-2} - b^{-2})}} e^{-N (\lambda b  - \log \sinh b + \log b)},
\eey
where $b$ is a function of $\lambda$, defined by the solution of the equation $\coth b - b^{-1} = \lambda$. At the limit $N\rightarrow \infty$, the results of the saddle-point approximation   coincide with the exact results.
Hence, the entropy $S$ is a sum of two terms 
\bey
S = S^{(0)} + \Delta S,
\eey
where $S^{(0)}$ is the entropy of the ideal gas in absence of an external field,
and
\bey
\Delta S(\lambda)  = - N \, [\lambda \, b(\lambda)  - \log (\sinh b(\lambda) / b(\lambda))]
\eey
is the contribution to the entropy from the gravitational field. We plot $\Delta S$ as a function of $\lambda$ in Fig. \ref{entropy1}. The entropy change $\Delta S$ is always negative,  and it is a concave function of $Q$.
The delta function with respect to $Q$ in the microcanonical distribution has decreased the size of the relevant phase space region.

We evaluate the gravitational acceleration $g = T \left(\partial S/\partial Q \right)_{U, N} = -2 T b/(mL)$, from which we obtain the equation  of state 
\bey
q = \frac{ T}{mg } - \frac{L}{2} \coth\left( \frac{mgL}{2T} \right). \label{qg0}
\eey
In the limiting cases $g\rightarrow 0$ and $g\rightarrow \infty$, $q \simeq (mL^2g)/(2T)$ and $q \simeq - \left(L/2\right) + T/(mg)$. In the former case, $q$ is proportional to $g$. In the latter case,
 the center of mass is at the bottom of the box;  the gas becomes effectively two-dimensional, as it was suggested in Ref. \cite{SaBa}.

The total entropy satisfies $S(tU, tN, t^{2/3}A, t^{1/3}L, t^{4/3}Q) = t S(U, N, L, Q)$.  In the long-box limit  $(L \rightarrow \infty)$, $\lambda$ is close to $-1$, and $\Delta S \simeq N \log(\lambda + 1)$. In the short-box limit ($L \rightarrow 0$), $\lambda$ is close to zero, and  $\Delta S  \simeq - \frac{3}{2} N \lambda^2$.

The physics of a column of gas in a gravitational field has long been understood. However, there has been little systematic study of the associated thermodynamic observables. As shown in the Appendix A, the changes in thermodynamics quantities due to gravity are functions of the dimensionless parameter $\epsilon = mgL/T$. For experiments on Earth, $\epsilon << 1$. In this regime,   the shift in the center of mass is $q = - 3  \, \epsilon L / 4$, the asymmetry $\Delta = \epsilon^{{} 2}/ 12$, and the change in the heat capacity per molecule $\Delta c_v = 3 \, \epsilon^2 / 4$. 

To estimate the feasibility of  measuring those quantities, we consider the heaviest noble gas (radon, with mass $m = 3.7\times 10^{-25}$ kg) at a temperature $T = 250^o$K well above its boiling point, and with  an extreme but feasible box height $L = 100$m. Then, $\epsilon \simeq 0.11$, and we find that $q = -8.2$m, $\Delta = 0.001$, and $\Delta c_v = 0.009$. The latter two quantities are within current accuracies of measurements of pressure and heat capacity.

\subsection{Canonical distribution}
An alternative approach to statistical mechanics is provided by Jaynes principle \cite{Jaynes}, according to which a statistical system is described by the distribution that maximizes the Shannon entropy 
\bey
S = - \int \frac{d^{3N}x d^{3N}p}{(2\pi \hbar)^{3N}N!} \rho(x, p) \log \rho(x, p), \nonumber
\eey
subject to appropriate constraints. For a box of gas, the  constraints are $U = \langle \hat{H}_0\rangle $ and $Q = \langle \sum_{i=1}^N m_i x_{3i}\rangle$. Then, entropy maximization yields the canonical distribution
\bey
\rho_{\beta, \eta}(x, p)  = \frac{ e^{-\beta H_0(p, q) - \eta  \sum_{i=1}^N m_i x_{3i}}}{\tilde{Z}(\beta, \eta, N) }, \label{candist}
\eey
where 
\bey
\tilde{Z}(\beta, \eta, N) = \int \frac{d^{3N}x d^{3N}p}{(2\pi \hbar)^{3N}N!} e^{-\beta H_0(p, q) - \eta  \sum_{i=1}^N m_i x_{3i}}.
\eey
Here, $\beta$ and $\eta$ are Lagrange multipliers. 

In this approach, the Shannon entropy for the entropy-maximizing probability distribution coincides with the thermodynamic entropy. We obtain $S = \beta U + \eta Q +\log \tilde{Z}$. This implies that
  $  \log \tilde{Z}$ is a Massieu function, namely, the double Legendre transform of the entropy with respect to $U$ and $Q$. We can therefore identify the Lagrange multipliers as   $\beta = (\partial S/\partial U)_{Q, N} = T^{-1}$ and $\eta =  (\partial S/\partial Q)_{U, N} = g/T$. 
 Hence, the probability distribution describes  a system in contact with a reservoir at temperature $\beta^{-1}$ and gravitational field $\eta/\beta$. 
 This means that     we can identify $-T \log \tilde{Z}$   with the Gibbs free energy $G(T, N, g) = E -TS $. For similar definitions in magnetic systems, see Ref. \cite{Castel}.

For an ideal gas, we find that $\tilde{Z}(\beta, \eta, N) =  Z_0(\beta, N) \zeta(\eta)$,
where $Z_0(\beta, N)$ is the partition function in absence of an external field, and
\bey
\zeta(\eta) = \left(\frac{\sinh(\eta m L/2)}{ \eta m L/2}    \right)^N
\eey
is the Laplace transform of $m\gamma(Q)$.

It is a standard result that a system in contact with a thermal reservoir at temperature $T = \beta^{-1}$ is described by    
the phase space distribution 
\bey
\rho_{\beta, Q}(x, p)  = \frac{1}{ Z (\beta, Q, N) } e^{-\beta H_0(p, q)} \delta(   \sum_{i=1}^n x_{3i} - Q/m) \delta({\bf P}),
\eey
where 
\bey
Z(\beta, N, Q)  \int \frac{d^{3N}x d^{3N}p}{(2\pi \hbar)^{3N}N!}  e^{-\beta H(x, p)} \delta(   \sum_{i=1}^n x_{3i} - Q/m)  \delta({\bf P})
\eey
is the Laplace transform of $\Gamma(U, N, Q)$ with respect to $U$.   The partition function is   identified with 
  $e^{-F(T, N, Q)/T}$, where $F$ is the Helmholtz free energy. For an ideal gas,
\bey
Z(\beta,  Q, N) = Z_0(\beta, N) \gamma(Q).
\eey
 
In the general case, equivalence between the different distributions is guaranteed if the successive Laplace transforms that connects $\Gamma$ to $Z$ to $\tilde Z$ are accurately evaluated by the saddle point method at the limit $N\rightarrow \infty$. The analysis of the Laplace transform from $\Gamma$ to $Z$ is standard textbook material \cite{Huang}. For the 
  Laplace transform connecting $Z$ to $\tilde{Z}$,
\bey
\tilde{Z}(\beta,  \eta, N) = \int dQ Z(\beta, N, Q) e^{-\eta Q} = \int dQ e^{-\beta F(\beta, N, Q) - \eta Q},
\eey
the saddle point approximation yields $\tilde{Z} = \sqrt{\pi N\chi_T/\beta} e^{-\beta G}$, hence, the distributions $\rho_{\beta, Q}, \rho_{\beta, \eta}$, and $\rho_{U, Q}$ are equivalent as long as the Helmholtz free energy $F$  scales with $N$, and $\chi_T$ does not vanish or diverge.
 
 Finally, we note the existence of two grand partition functions $\tilde{\Xi}(z, \beta, \eta) = \sum_{N=0}^{\infty} z^n \tilde{Z}(\beta, \eta, N)$ and $ \Xi (z, \beta, Q) = \sum_{N=0}^{\infty} z^n  Z(\beta, Q, N)$, where $z = e^{\beta \mu}$. The Landau potential is defined as $\Phi = -T \log \Xi$ and the transformed Landau potential as $\tilde{\Phi} = -T \log \tilde{\Xi}$.

\section{A self-gravitating system}

In this section, we show that our   thermodynamic analysis extends to self-gravitating systems. 
To this end, we consider 
the simplest example: a column (one-dimensional box  of height $L$)
 of self-gravitating ideal gas. In particular, we show that the gravitational pull arises naturally as a quantity in the thermodynamic fundamental space associated to an external field. It enables a distinction between the external gravitational field and the total gravitational field in the system, analogous to the distinction between the electric field ${\bf E}$ and the electric displacement ${\bf D}$ in dielectrics.
 
 One-dimensional models of self-gravitating systems have been widely studied, both in the equilibrium and the non-equilibrium context, because they admit exact solutions and they can be generalized for relativistic systems---see, for example, Refs. \cite{Salz, Rybicki, Mann1, Mann2}. At the limit of large particle numbers and in the mean-field approximation, the thermodynamic properties of self-gravitating systems can be defined in terms of hydrodynamics in local equilibrium. The condition of local equilibrium means that the fluid is well described by an entropy density $s(x)$, which is a local functional of free energy density $u(x)$ and the particle densities $n_a(x)$, where $a$ stands  for the different particle species. As shown in Sec. 2.2, the temperature $T$ is constant in the fluid, so we are interested in the values of the entropy density $s(x)$ along isotherms.
 
 Gravitational interactions are described in terms of 
 the gravitational potential $\phi$ that satisfies Poisson's equation $\nabla^2\phi = 4 \pi G \rho$. The mass density is defined as $\rho(x) = \sum m_a n_a(x)$, and the  pressure satisfies the hydrostatic equation  $\nabla P = - \rho \nabla \phi$. This set of equations can be solved, and the thermodynamics of the system is determined by the entropy  $S = \int dx s(x)$. The entropy $S$ is a function of   the boundary variables and the conserved quantities. 
 
 The simplest case corresponds to an ideal gas with a single particle species. The entropy density is $s = (\rho/{m}) \, \log(b \, t^{3/2}/\rho)$, where $b$ is a constant and $t = T/m$. The associated equations of state are $u = 3 \, t \rho/2$ and $P = t \rho$. The spherical symmetric solution of the hydrodynamic equations  in three dimensions is the {\em isothermal sphere}, a system that has been extensively analyzed \cite{BiTr, Chandr}. Here, we restrict to one spatial dimension, so that there is a straightforward  correspondence with the results of the previous sections. Note that our conclusions in this section apply to any equation of state, subject to the constraint of local equilibrium. The assumption of an ideal gas enables a fully analytic treatment.
 
 The equation of hydrostatic equilibrium in one dimension gives $t \rho' = - \rho \,\phi'$, which implies
   that $(t \log \rho + \phi)' = 0$, with solution $\rho = \rho_0 e^{-\phi/t}$, 
 where $\rho_0$ is an integration constant. Substituting into Poisson's equation, we obtain  
 \bey
 \phi'' = 4\pi G \rho_0 e^{-\phi/t}.
 \eey
 This is analogous to Newton's equation for a potential $V(\phi) = 4\pi G\rho_0 t e^{-\phi/t}$. Its solution is 
 \bey
e^{\phi/t} = \frac{8\pi G \rho_0}{k^2t} \cosh^2[k(x-x_0)], ,
 \eey
 where $x_0$ and $k$ are   integration constants. The mass density reads
 \bey
 \rho(x) = \frac{k^2 t}{8\pi G \cosh^2[k(x-x_0)]}.
 \eey
 Of the three integration constants, $\rho_0$ corresponds to the arbitrary choice of zero for the potential, and it has no thermodynamic significance; $k$ and $x_0$ correspond to the total particle number $N$ and the gravitational pull $Q$, respectively. To see this, we evaluate 
 \bey
  N &=& m^{-1} \int_{-L/2}^{L/2} dx \rho(x) = \frac{kt}{8 \pi Gm}\left[\tanh[k(x_0+L/2)] - \tanh[k(x_0 -L/2)]\right]
\\
  Q &=&   \int_{-L/2}^{L/2} dx x \rho(x) =  Nm x_0.
 \eey
 Let  $y(x, \lambda) = x [\tanh[x(1+\lambda)] + \tanh[x(1-\lambda)]]$, and $w(y, \lambda)$ be its inverse with respect to $x$. Then, we can solve for the integration constants,
  \bey
k = \frac{2}{L} w(\dfrac{4 \pi G m NL}{t}, \lambda), \hspace{1cm} x_0 = q,
 \eey
 where $\lambda = 2q/L$. 
 
We evaluate the entropy
 \bey
 S = \int_{-L/2}^{L/2} dx s(x) =  N \log \left(\frac{8\pi G b t^{1/2}}{k^2}\right) + \frac{kt}{4\pi G m} \left[F(kL/2+kx_0) - F(-kL/2+kx_0)\right],
 \eey
  where $F(x) = \tanh x( 1 + \log \cosh x) - x$.
 
\begin{figure}[tb] 
\includegraphics[height=5cm]{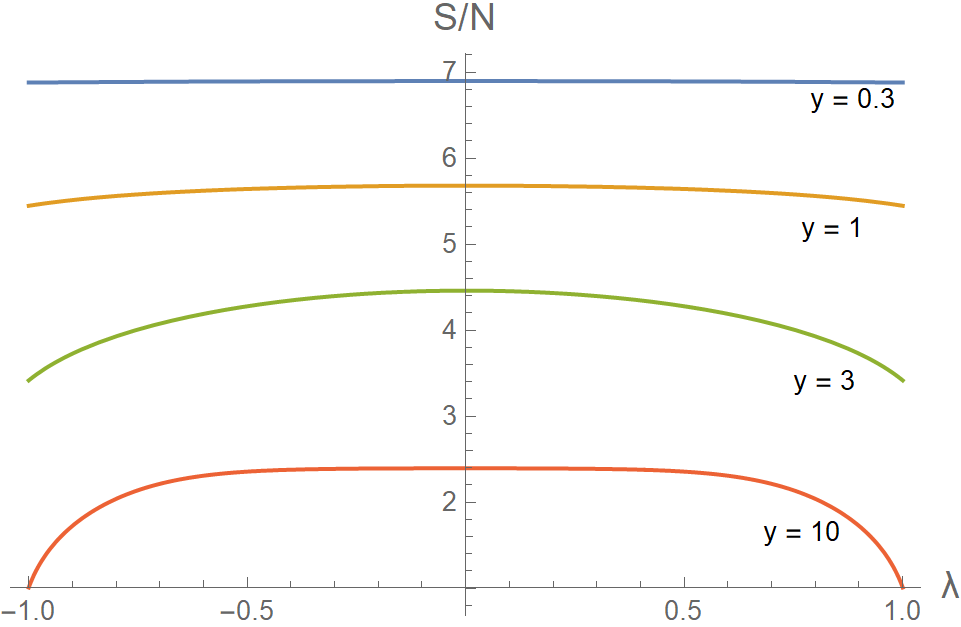} \caption{We plot the entropy per particle $ S/N$   as a function of the dimensionless gravitational pull $\lambda$ and for different values of $y$.}
\label{entsas}
\end{figure}

 We see that the fundamental space indeed consists of the variables $U, N, L$, and $Q$, where $U = \frac{3}{2} NT$---in agreement with the analysis of Sec. 2. That is, the fundamental thermodynamic space for this system coincides with the space of parameters that characterize solutions to the constitutive equations, namely, the hydrostatic equation and the Poisson equation). We can write the entropy explicitly as 
 \bey
 S &=& N \log(2\pi \, G \,  b \, t^{1/2}L^2) -2N \log w(y, \lambda) \nonumber \\
 &+& \frac{2 N }{y}w(y, \lambda) \left[F[ w(y, \lambda)(1+\lambda)] + F[ w(y, \lambda)(1-\lambda)] \right]
 \eey
where $y =  4 \pi G m LN /t $. 
 The entropy is invariant under the transformation $\lambda \rightarrow -\lambda$; for fixed $y$, it is maximized at $q = 0$---see Fig. \ref{entsas}. In general, the entropy decreases with $y$. We also note the rescaling property
 \bey
 S(tN, t^{9/5}U, t^{-1/5}L, t^{6/5}Q) = t S(N, U, L, Q). \nonumber
 \eey
 
 To understand the relation of thermodynamic quantities with the gravitational field, we first recall that the external gravitational field $g_{ext}$ can be read from the derivative of the entropy with respect to $Q$, 
 $g_{ext} = \partial S/\partial Q = (2t/NL) \partial S/\partial \lambda$. 
 
 On the other hand, the gravitational field is $g(x) = \phi'(x) = 2tk \tanh(k(x-x_0))$. We write $g = \phi'$ rather than $g = -\phi'$, to agree with the convention of Sec 2.2 that $g$ points downwards. The potential outside the box is a solution of the Poisson equation for the vacuum, i.e., it is of the form $\phi = g x + c$. By continuity,  the fields $g_{\pm}$ above (+) and below (-) the box are:
  $g_{\pm} = \pm 2tk \tanh[kL/2(1\mp \lambda)]$. By Poisson's equation the difference $g_+ - g_- = 4\pi G m N$ is determined by the number of particles in the box, and it is independent of the gravitational pull. Hence, it is the average external field $\bar{g} = \frac{1}{2}(g_+ + g_-)$ that carries the $q$-dependence.

\begin{figure}[tb] 
\includegraphics[height=5cm]{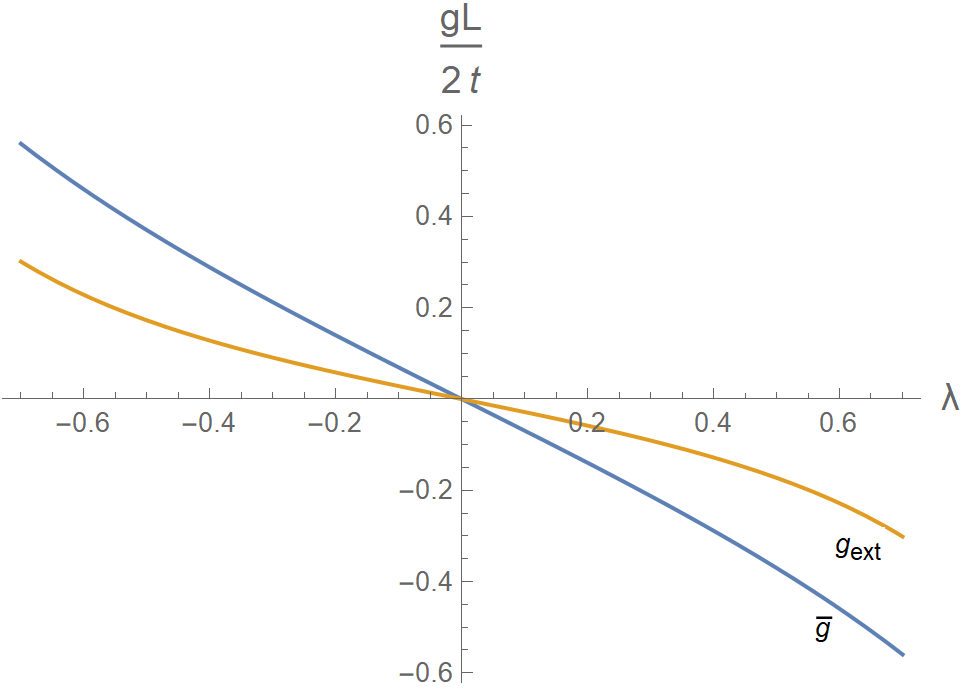} \caption{We plot the dimensionless versions of the external field $g_{ext}$    and the average total field $\bar{g}$   (multiplied by $\frac{L}{2t}$) that characterize the system, as functions of the dimensionless gravitational pull $\lambda = 2q/L$. In this plot, $y = 1$.}
\label{fourfields}
\end{figure}

\begin{figure}[tb] 
\includegraphics[height=5cm]{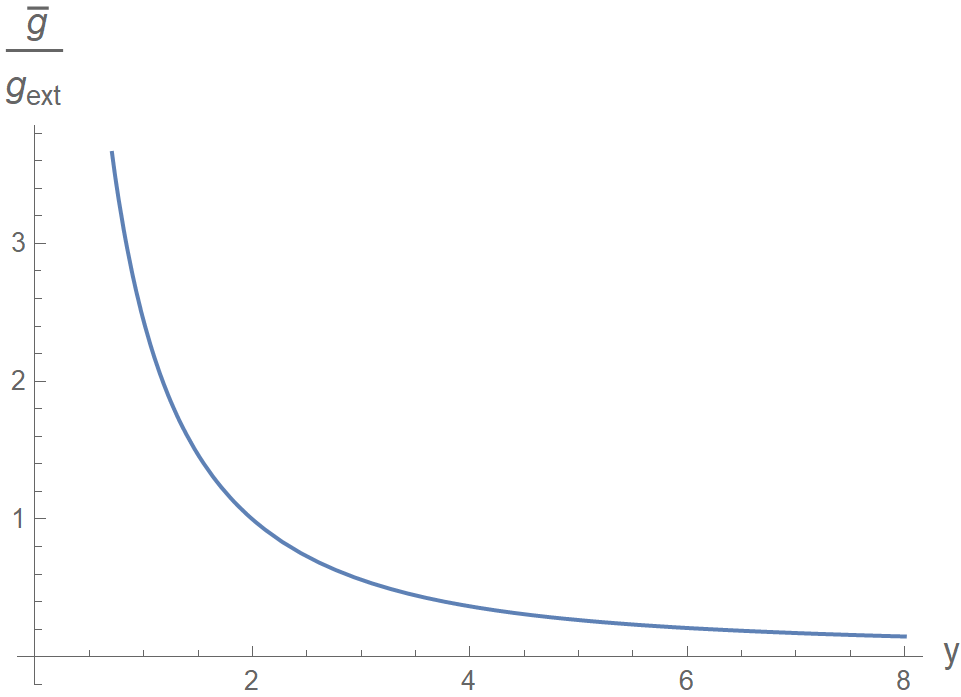} \caption{We plot the ratio $\bar{g}/g_{ext}$   as  a function  of the parameter $y =  4 \pi G m LN/t$,   for $\lambda = 0.1$.  }
\label{response}
\end{figure}

 The field $\bar{g}$ does not coincide with the external field $g_{ext}$---see Fig. \ref{fourfields}. The field  $g_{ext}$ is the external field in which the box is initially placed, while the field  $\bar{g}$   incorporates the system's self-gravity. The average field is not always larger than the external field. As shown in Fig. \ref{response}, the ratio $\bar{g}/g_{ext}$ drops with $y$, and becomes very small as $y \rightarrow \infty$.
 
   When comparing with dielectric or paramagnetic systems, the   average field $\bar{g}$ corresponds to the electric field ${\bf E}$ and the magnetic field ${\bf B}$, respectively; the external field $g_{ext}$ corresponds to the electric displacement ${\bf D}$ and the magnetic   intensity ${\bf H}$. Obviously, this analogy does not go very far. First, gravity is always attractive, and there is no way to screen it. Second, gravity is a long-range force, so the difference between  $\bar{g}$ and $g_{ext}$  persists even outside matter. Nonetheless, it is tempting to interpret the regime $\bar{g}/g_{ext} > 1$ as analogous to paramagnetism and the regime  $\bar{g}/g_{ext} < 1$ as analogous to diamagnetism. The gravitational analogue of ferromagnetism would be a system in which the vanishing of the gravitational pull does not correspond to an entropy maximum, as in Fig. \ref{entsas}.

 
We evaluate the   potential difference at the box boundary, 
 \bey
 \Delta \phi &=& \phi(L/2) - \phi(-L/2) = 2 t \log\left[ \frac{\cosh( \frac{1}{2}kL (1-\lambda))}{\cosh( \frac{1}{2}kL (1 + \lambda))}\right].
 \eey
 We note that $\Delta \phi$ in an increasing function of $|\lambda|$. Since $P_t/P_b = \rho(L/2)/\rho(-L/2) = e^{-\Delta \phi/t}$, we see that  $q$ effectively measures the difference in pressures between top and bottom of the box, as expressed in the barometric formula.

\section{Inhomogeneous field}
In this section, we consider  thermodynamics in presence of an external non-homogeneous  gravitational field $\phi({\bf x})$. Let $\rho({\bf x})$ be the mass density of the 
system. A change $\delta \rho({\bf x})$ in the mass density corresponds to work, $dW = - \int d^3x \phi({\bf x}) \delta \rho({\bf x)}$ gained by the system. 
Hence, we obtain a version of the first law: $dU = TdS - \int d^3x \phi({\bf x}) \delta \rho({\bf x})$. For simplicity, we ignored the pressure and chemical potential terms. 
The main ideas in the analysis of Sec. 2 apply here, modulo some technical modifications. 

Let the 
   background field be generated by masses $M_k$ each located at fixed positions ${\bf R}_k$.  The generalization to a continuous mass distribution as sources of the background gravitational field is straightforward. Then, $\phi({\bf x}) = - \sum_k GM_k/|{\bf x} - {\bf R}_k|$, where $G$ is Newton's constant. The mass density   is expressed in terms of the particles' coordinates as 
$\rho({\bf x}) = \sum_{i=1}^N m_i \delta({\bf x} - {\bf x}_i)$. The work term becomes $dW = - \sum_k GM_k \delta Q_k$, where
\bey
Q_k = \sum_{i=1}^N\frac{m_i}{|{\bf x}_i - {\bf R}_k|}, \label{qkin}
\eey
is the gravitational pull generated by the mass $ M_k$. 
The gravitational pull $Q_k$ and the associated mass $GM_k$ form a conjugate thermodynamic pair. 

To keep the units of the gravitational pull the same as in the previous sections, we can multiply $Q_k$ with the square of an arbitrary length $r_k$. We divide accordingly the conjugate variable $M_k$. Note that for a mass $M_k$ localized within a sphere of radius $r_k$, the quantity $GM_k/r_k^2$ is the gravitational acceleration at the sphere. In what follows, we will work with Eq. (\ref{qkin}).

The microcanonical distribution becomes
\bey
\rho_{U, Q} (x, p) = \Gamma(U, N)^{-1} \delta(H_0 - U) \prod_k\delta(   \sum_{i=1}^n |{\bf x}_i - {\bf R}_k|^{-1}  - Q_a/m) \delta({\bf P}), \label{microuq3}
\eey
where 
\bey
\Gamma(U, N, Q) = \int \frac{d^{3N}x d^{3N}p}{(2\pi \hbar)^{3N}N!}   \delta(H_0  - U) \prod_k \delta(   \sum_{i=1}^n |{\bf x}_i - {\bf R}_k|^{-1}  - Q_a/m)  \delta({\bf P}).
\eey
As an example, we consider the case of an ideal gas contained in a spherical cavity of radius $R$, at the center of which lies an external mass $M$. Then, $\phi({\bf x}) = - GM/|{\bf x}|$, and $Q = m\sum_{i=1}^N |{\bf x}_i|^{-1}$. 

It is straightforward to show that $\Gamma(U, N, R, Q) = \Gamma_0(N, U, V) \gamma(Q)$, where $\Gamma_0(N, U, V) $ is the volume of the energy surface  for  a gas in volume $V = \frac{4}{3}\pi R^3$, and 
\bey
\gamma(Q) = \frac{1}{V^N} \int d^{3N}x \delta(    \sum_{i=1}^n |{\bf x} |^{-1}  - Q/m).
\eey
For calculational purposes, it is convenient to work with the analogue of the canonical distribution (\ref{candist}). The partition function factorizes as $\tilde{Z}(N, \beta, R, GM) = \tilde{Z}_0(N, \beta, V) \zeta(\beta GM )$, where $\tilde{Z}_0(N, \beta, V)$ is the partition function of an ideal gas in a volume $V = 4 \,\pi R^3 /3 $. The function   $\zeta $ is the Laplace transform of $m\gamma(Q)$,
\bey
\zeta(\eta) = \left[\frac{1}{V} \int_{r<R} d^3 x e^{- \eta m/r}   \right]^N = \left[\tau(\eta m/R)\right]^N,
\eey
where   $\tau(x) = \frac{1}{2}\left(e^{-x}(2-x + x^2) + x^3 \mbox{Ei}(x)   \right)$, and Ei is the exponential integral function

It follows that the Gibbs function is $G(N, T, R, GM) = F_0(N, T, R) - N T\log \tau ( GMm/(RT))$, where $F_0$ is the Helmholtz free energy of the ideal gas.  We compute the gravitational pull
\bey
Q  =  T \frac{\partial \log \zeta}{\partial (GM)} =  NmR^{-1} {\cal B} \left(\frac{GM}{TR}\right),
\eey
 where ${\cal B} (x)= - (\log \tau(x))'$. The function ${\cal B} (x)$  satisfies ${\cal B}(0) = \dfrac{3}{2}$, and it decreases monotonically towards an asymptotic value $1$.

The fact that the gravity contribution depends only on the combination $GM/(RT)$ implies   relations  between the gravitational contribution to entropy $\Delta S $, the gravitational contribution $\Delta P_r$  to the radial pressure at the boundary, and $Q$,
\bey
\Delta P_r &=&  \frac{T}{4\pi R^2} \frac{\partial \log \zeta}{\partial R} = \frac{GM}{4\pi R^3}Q \\
\Delta S &=& \frac{\partial (T \log \zeta)}{\partial T} = N \log \tau  \left(\frac{GMm}{TR}\right) + \frac{GMQ}{T}.
\eey

\section{Quantum gases}
In this section, we   generalize the results of Sec. 4 to quantum gases. In particular, we show that there exists a continuous phase transition for fermions at zero temperature.
\subsection{The equilibrium density matrices}
For a quantum gas  in a homogeneous gravitational field ${\bf g} = (0, 0, g)$, the internal energy $U$ is obtained from the Hamiltonian operator $\hat{H}_0 = \sum_{i=1}^N ({\hat{\bf p}_i^2}/{2m_i}) + \sum_{i=1}^N \sum_{j < i}  V(\hat{\bf x}_i - \hat{\bf x}_j) $. We also define the gravitational-pull operator $\hat{Q} = \sum_{i=1}^N m_i \hat{x}_{3i}$, and the center-of-mass momentum $\hat{\bf P}  = \sum_i \hat{\bf p}_i$.

We find the commutation relations 
\begin{align}
[\hat{H}_0, \hat{\bf P}] = 0,\hspace{1cm}
[\hat{Q}, \hat{H}_0] = \hat{P}_3, \hspace{1cm}
[\hat{Q}, \hat{P}_3] = i \sum_{i=1}^N m_i \hat{I}. \nonumber
\end{align}
We construct the micro-canonical representation in the subspace of zero center-of-mass momentum, where  $[\hat{Q}, \hat{H}_0] = 0$. We define the microcanonical density matrix
\bey
\hat{\rho}_{mic} = \frac{1}{\Gamma(U, Q, N)} \delta(\hat{\bf P}) \delta(\hat{H}_0 - U) \delta(\hat{Q} - Q)\delta(\hat{\bf P}),
\eey
where $\Gamma(U, Q, N) = Tr [\delta(\hat{\bf P}) \delta(\hat{H}_0 - U) \delta(\hat{Q} - Q)\delta(\hat{\bf P})]$ is the energy volume.

The double Laplace transform of $\Gamma$ is  
\bey
\tilde{Z}(\beta, \eta, N) = \int_0^{\infty} dU \int_0^{\infty} dQ \Gamma(U, Q, N) e^{-\beta U - \eta Q} = Tr_{\hat{\bf P} = 0} (e^{-\beta \hat{H}_0} e^{-\eta \hat{Q}}) = Tr_{\hat{\bf P} = 0} (e^{-\beta \hat{H}_0 -\eta \hat{Q}}),
\eey
In the subspace  $\hat{\bf P} = 0$, $\tilde{Z}$ is constructed from the eigenvalues of the total Hamiltonian $\hat{H} = \hat{H}_0 + g \, \hat{Q}$, where $g = \eta/\beta$.  

For quantum ideal gases, both bosonic and fermionic, it is convenient to employ  the grand canonical ensemble. In this case, thermodynamic quantities are constructed from the single-particle density of states $\rho(E)$ of $\hat{H}$. The calculation is straightforward---see the Appendix B.
\bey
\rho(E) = \frac{2\sqrt{2}A}{3 \pi^2 \sqrt{m}g} \times\left\{ \begin{array}{cc} \left[\left(E - mg \ell_b \right)^{3/2} - \left(E - mg \ell_t\right)^{3/2}\right], & E > mg \ell_t \\ \left(E - mg \ell_b \right)^{3/2}, & mg \ell_b \leq E\leq  mg \ell_t \end{array} \right. \label{rhoe}
\eey
The branch for $E < mg \ell_t$ corresponds to classical orbits in which the particle does not reach the top of the box. The branch for $E\geq mg\ell_t$ corresponds to classical orbits in which the particle reaches the top of the box, and it is reflected elastically.

\subsection{A phase transition for fermions}

As an example, we calculate the zero-temperature thermodynamics of a gas with $N$ fermions (spin $\frac{1}{2}$) in a gravitational field---for past work, see Ref. \cite{Kole, Thai}. It is convenient to choose coordinates so that $\ell_b = 0$ and $\ell_t = L$. At zero temperature, the Gibbs free energy coincides with the total energy $E = \int_0^{\epsilon_F} dE \rho(E) E $; the Fermi energy $\epsilon_F$ is defined by 
 $\int_{0}^{\epsilon_F} dE \rho(E) = N$.  
 
 We find that 
\bey
N = \frac{c_0 Am^2g^{3/2} L^{5/2}}{5} \, f_{1}(x), \hspace{1cm}
E = \frac{c_0 A m^3 g^{5/2} L^{7/2}}{7 } \,  f_{2}(x),
\eey
 where 
 $x = \epsilon_F/(mgL)$,  $c_0 = 4\sqrt{2}/(3\pi^2) \simeq 0.19$, and
 \bey
 f_{1}(x) &=& \left\{ \begin{array}{cc} x^{5/2} - (x - 1)^{5/2}, & x \geq 1, \\[9pt] x^{5/2}, & x < 1,\end{array} \right.
\\[15pt]
 f_{2}(x) &=& \left\{ \begin{array}{cc} x^{7/2} - (x - 1)^{7/2} - \frac{7}{5} \,(x-1)^{5/2} \, , & x \geq 1, \\[9pt] x^{7/2}  , & x < 1.\end{array}\right.
 \eey
 
 We obtain,
 \bey
 E = \frac{c_0 A m^3 g^{5/2} L^{7/2}}{7 } \  \sigma \left(\frac{5N}{c_0 A m^2 g^{3/2} L^{5/2}}\right),
 \eey
 where $\sigma(x) = f_{2}[f^{-1}_{1}(x)]$. The function $\sigma$ is continuous, and it has continuous first and second derivatives at $x = 1$. However, its third derivative at $x = 1$ is discontinuous. This implies that the system is characterized by a continuous phase transition. 
 
For $x < 1$, $\sigma(x) = x^{7/5}$, and we find 
\bey
E = \frac{5}{7} (5/c_0)^{7/5} \frac{N^{7/5}m^{1/5}g^{2/5}}{A^{2/5}}.
\eey
The total energy $E$ does not depend on $L$. This phase describes a fluid that does not reach the top of the box. 
 The energy per particle $E/N$ is proportional to $(N/A)^{2/5}$, i.e., the system behaves  effectively as two dimensional. The gravitational pulls $Q = (2/5)(E/g)$, which implies that $U = E - Qg = (3 /5) \ E$.
The pressure at the top of the box is zero, and at the bottom equals $Mg/A$. 
 The horizontal pressure  is $P_h = (2/5) (E/A)$. 
 

For $x >> 1$, $\sigma(x) \sim x^{5/3}$. In this case,  the energy is given by the standard $g$-independent expression for degenerate fermions. As $g$ increases, the pressure asymmetry increases, until a phase transition occurs at $g = g_c$, where 
$g_c =  \dfrac{(5 N/c_{0})^{2/3}}{A^{2/3}m^{4/3}L^{5/3}}$. 
Equivalently, we can keep $g$ constant  and vary $L$. Then,  the transition occurs at 
$L = L_c$, where 
\bey
L_c = \left(\frac{5N}{c_0 A m^2 g^{3/2}}\right)^{2/5}.
\eey
Then, 
\bey
E = \frac{5}{7}\, N \, (mg L_c) \, (L/L_c)^{7/2} \,  \sigma[{} \,(L_c/L)^{5/2}].
\eey

In the vicinity of $x = 1$, 
\bey
\sigma(x) \simeq x^{7/5} + \frac{14}{15}\sqrt{\frac{2}{3}} (x-1)^{5/2} \, \Theta(x-1),
\eey
where $\Theta$ is the  step function. Hence, for $L$ near the critical value $L_c$, we can express the vertical pressure $P_v$ as 
\bey
P_v = \frac{125}{36} \sqrt{\frac{5}{3}} \frac{Mg}{A} \left(1 - \frac{L}{L_c}\right)^{3/2} \Theta(L_c - L).
\eey
 The second derivative of the pressure is discontinuous at $L = L_c$.
  
It is surprising that the seemingly innocuous property of the quantum gas not being able to reach the top of the box is manifested as a phase transition, 
i.e., in a non-smooth state function. This phenomenon may have
non-trivial physical implications. The height $L$ of the box cannot be made arbitrarily small, there is a minimum value $L_0$ that corresponds to the coarse-graining necessary for the gas to define a thermodynamic system. If we take a box of size $L_0$ as an elementary volume of the fluid in local equilibrium, a strong field such that $L_c < L_0$ will affect the conditions of local equilibrium. The result  will be a local equation of state that depends on the background field. Such a behavior is thermodynamically consistent,  because the gravitational field is a thermodynamic variable. To check this hypothesis, we need to study how the phase transition is modified by the presence of interactions, and also to 
analyze the quantum stress-energy tensor inside the box.


 



\section{Relativistic acceleration}

In this section, we generalize the analysis of previous sections to relativistic systems. In particular, we consider a gas of particles in a static gravitational field. We focus  on the case of constant proper acceleration (Rindler spacetime)---see Refs.  \cite{SouAn23, LMar, Pad2, Pad3} for past works on this topic---but our results straightforwardly generalize to inhomogeneous fields. 

We consider static spacetime geometries of the form
\bey
ds^2 = -C(x)^2 dt^2 + \delta_{ij}dx^i dx^j, \label{metric}
\eey
where $x_i$ are spatial coordinates and $C(x)$ is the lapse function. For Rindler spacetime with proper acceleration $g$ along the axis 1, $C(x) = 1 + {\bf g} \cdot {\bf x}$.

The Hamiltonian for a particle of mass $m$ in a metric (\ref{metric}) is given by $H(x, p) = C(x) H_0(p)$, where $H_0(p) = \sqrt{{\bf p}^2 + m^2}$. The total Hamiltonian of  system of $N$ particles is given by $H({\bf x}_i, {\bf p}_i) = \sum_{i=1}^N C(x_i) H_0({\bf p}_i)$, where ${\bf x}_i$ is the position  and ${\bf p}_i$ is the momentum of the $i$-th particle. We note that $\partial H/\partial {\bf g} = \sum_i H_0({\bf p}_i)  {\bf x}_i$, so the thermodynamic conjugate to the proper acceleration ${\bf g}$ is 
\bey
{\bf Q} = \sum_i H_0({\bf p}_i) {\bf x}_i.
\eey
  This quantity is the relativistic version of the gravitational pull (\ref{Q}), the center of energy rather than the center of mass being the relevant relativistic quantity. 

Consistent relativistic interactions require quantum field theory, and for this reason, we only consider an ideal gas in this section---see Ref. \cite{SouAn23} for a treatment in terms of quantum fields. Again, we identify the internal energy $U$ with the total kinetic energy. We assume the same setup as in the non-relativistic case: a box of $N$ particles, with area $A$ normal to the acceleration, the bottom at $x_3 = \ell_b$ and the top at $x_3 = \ell_t$.   
The fundamental thermodynamic space $\Lambda$ consists of the variables $N$, $U$, $A$, $\ell_t$, $\ell_b$, and $Q$. The entropy $S$ is a function on $\Lambda$. The first law of thermodynamics reads
\bey
dU = TdS - P_h L dA + P_b A d\ell_b - P_t A d\ell_t + \mu d N - g d Q. 
 \label{rep1vv}
\eey

Space translation acts on $\Lambda$ as: $\ell_t \rightarrow \ell_t + a, \ell_b \rightarrow \ell_b + a, Q \rightarrow Q + U a$. Unlike the non-relativistic case, $Q$ is transformed   by a term proportional to on the internal energy $U$. Invariance of the action under space translation implies the relation $P_b - P_t = Ug/A$. Hence we can substitute the term  $P_b A d\ell_b - P_t A d\ell_t$ with $-P_v dL$, where $L = \ell_t - \ell_b$, and 
\bey
P_t = P_v - \frac{Ug}{2A}, \hspace{1cm} P_b = P_v + \frac{Ug}{2A}.
\eey
We follow the arguments of Sec. 2.2., in order to find how   temperature varies inside the box. We maximize the entropy $S = \int d^3 x \, s[\rho(x), n(x)]$ for constant particle number $N = \int d^3x n(x)$, internal energy $U = \int d^3x \, \rho(x)$, and gravitational pull $Q = \int d^3x \,  x \rho(x)$. We find that
\bey
\partial s/\partial \rho = \beta + \eta x, \hspace{1cm} \partial s/\partial n = \gamma,
\eey
where $\beta, \gamma$, and $\eta$ are Lagrange multipliers. This means that $T(x)  (1 + \eta/\beta x) = \beta^{-1}$, and $\mu(x)/T(x) = \gamma$. By identifying $\eta/\beta$ with $g$, we obtain {\em Tolman's law} $T(x) C(x) = \beta^{-1}$. The presence of the gravitational pull in the fundamental space is essential in order for Tolman's law to be compatible with the maximum entropy principle.

The statistical mechanical analysis of Sec. 3 remains the same, modulo the change in the definition of the gravitational pull $Q$, and Tolman's law. 
The microcanonical distribution is 
\bey
\rho_{U, Q} (x, p) = \Gamma(U, N, Q)^{-1} \delta(H_0 - U) \delta(   \sum_{i=1}^n H_0({\bf p}_i) x_{3i} - Q) \delta({\bf P}), \label{microuqrel}
\eey
where 
\bey
\Gamma(U, N, Q) = \int \frac{d^{3N}x d^{3N}p}{(2\pi \hbar)^{3N}N!}   \delta(H_0  - U) \delta( \sum_{i=1}^n H_0({\bf p}_i)   x_{3i} - Q)  \delta({\bf P}).
\eey
The microcanonical entropy is defined as  $S(U, N , Q) = \log \Gamma(U, N, Q)$.

 For explicit calculations, it is convenient to work with the 
  Laplace transform $Z(\beta, N,  \ell_t, \ell_b, \eta)$ of $\Gamma(U, N, Q)$, which defines the Gibbs representation. Inserting back the dependence on the variables $A, \ell_t, \ell_b$,   invariance under space translation implies that
  \bey
 Z(\beta - a \eta/\beta, N, A, \ell_t + a, \ell_b + a, \eta) = Z(\beta, N, A, \ell_t, \ell_b, \eta). \nonumber 
  \eey
For an ideal gas,   
  $Z(\beta, N, \eta) = Z_1(\beta, \eta)^N/N!$, where
\bey
Z_1(\beta,  \eta) = \int \frac{d^{3}x d^{3}p}{(2\pi)^{3}}e^{-\beta H_0({\bf p}) - \eta H_0({\bf p}) x}.
\eey
We calculate 
\bey
Z_1(\beta, \eta) = \left\{ \begin{array}{cc}   \dfrac{Am^2}{\pi^2 \eta  } \, \left( \dfrac{K_1(\beta m + \eta m \ell_b)}{\beta m + \eta m \ell_b} - \dfrac{K_1(\beta m + \eta m \ell_t)}{\beta m + \eta m \ell_t}\right),& m \neq 0 \\[18pt]
\dfrac{A}{2\pi^2 
\eta  } \, \left( \dfrac{1}{(\beta + \eta \ell_b)^2} - \dfrac{}{}\dfrac{1}{(\beta + \eta \ell_t)^2} \right), & m = 0 \end{array} \right. \label{pfz0}
\eey 
 
In Ref. \cite{SouAn23}, we calculated the partition function  for a photon gas
\bey
\log Z =  \frac{2\pi^2A }{45\eta} \,  \left(\frac{1}{(\beta + \eta \ell_b)^2} - \frac{1}{(\beta + \eta \ell_t)^2} \right).  \label{photon}
\eey
In the Appendix C, we describe the thermodynamic properties of classical and quantum ideal gases of massless particles. 

\section{Conclusions}
We described our motivation, our longer term program, and our results in the Introduction. Here, we want to focus on the implications   of our results.

Since the gravitational pull remains a thermodynamic variable for self-gravitating systems (at least for models of level 2), the appropriate microcanonical distribution for such systems is given by Eq. (\ref{microuq}), or by its relativistic generalization (\ref{microuqrel}). These distributions differ from those that have been considered so far, and they suggest a more intricate relation between the canonical and microcanonical descriptions of such systems. 

For self-gravitating systems at level 3,  the determination of the fundamental space $\Lambda$ is a challenge. Ref. \cite{AnSav14} suggests that $\Lambda$ consists only of geometric properties of the enclosing boundary (metric and extrinsic curvature) and the numbers of particles of each species. However, a thermodynamic interpretation of the geometric variables is lacking, as is a physical prescription to determine the internal energy.  

The hypothesis of a generalized second law in black hole thermodynamics led Bekenstein to the proposal of an entropy bound \cite{BBB, BGSL}. For systems with negligible self-gravity, the entropy-to-energy ratio is bounded by the largest dimension $D$ of the system, $S/E \leq 2 \pi D$.
Although many systems are in good agreement with the entropy bound, there are counterexamples.  Ref. \cite{SouAn23} suggests that the entropy bound is violated in the presence of strong background gravitational fields. However, in this context, the distinction between the internal and total energy is crucial for the very definition of the entropy bound. In our opinion, entropy bounds should be derivable from first principles in an axiomatic framework for gravitational thermodynamics.

The phase transition of degenerate fermions in Sec. 6 provides an intriguing prospect. Although further work is needed in order to make it more concrete,  it suggests that strong gravitational fields modify 
 the equation of state for nuclear matter.  If true, it would influence the solutions to the equations of hydrostatic equilibrium (e.g., the Tolman-Oppenheimer-Volkoff equation), thus impacting the physics of compact stars.

\begin{appendix}

\section{Thermodynamic quantities for a gas in a homogeneous gravitational field}

\renewcommand{\theequation}{A-\arabic{equation}}
  \setcounter{equation}{0}

First,  we write Eq. (\ref{qg0}) as
\bey
q   = \frac{L}{2} \, \left( \frac{g_0}{g} - \coth(g/g_0)\right),. \label{qg0b}
\eey
where $g_0 = 2 \, T/(mL)$.

From Eq. (\ref{qg0b}), we evaluate the gravitational susceptibility
\bey
\chi_T = \frac{mL}{2g_0} \, \Big{(}(g_0/g)^2 - \csch^2(g/g_0)\Big{)}.
\eey
 We see that $\chi_T$ starts from a maximum value $\chi_T(0) = \dfrac{mL}{6g_0}$ and drops, decaying as $\chi_T \sim \dfrac{mLg_0}{2g^2}$ for $g >> g_0$.

Next, we evaluate the pressures
 
\bey
P_h = \frac{NT}{V},\hspace{1cm} P_v = \frac{Mg}{2A} \coth\left( \frac{Lmg}{2T} \right),
\eey
from which we obtain
\bey
\frac{P_t}{P_b} = \frac{ P_v - \resizebox{0.6cm}{0.45cm}{$\frac{Mg}{2A}$}}{P_v + \resizebox{0.6cm}{0.45cm}{$\frac{Mg}{2A}$}} = e^{-mgL/T},
\eey
 in accordance with the barometric formula. To the best of our knowledge, a proof of the barometric formula from the microcanonical distribution has been missing \cite{barometric}. This proof has the added benefit that it can be applied to both quantum and relativistic systems.

The associated compressibilities are
\bey
\kappa_{T}^{vv} =  \frac{4AT\sinh^2\left(\frac{mgL}{2T}\right)}{Nm^2 g^2 L}, \hspace{1cm}
\kappa_{T}^{vh} = \frac{1}{P_v}, \hspace{1cm}  
\kappa_{T}^{hv} = \kappa_{T}^{hh}  =  \frac{1}{P_h}.
\eey

Finally, we note that the heat capacity $c_V = \frac{T}{N} \left(\partial S/\partial T\right)_{A, L, g}$ 
\bey
c_V = c_V^{(0)} + mgq/T,
\eey
where $c_V^{(0)} = \frac{3}{2}$ is the heat capacity in absence of gravity,

\section{Calculating the single particle density of states, Eq. \ref{rhoe}}
 \renewcommand{\theequation}{B-\arabic{equation}}
  \setcounter{equation}{0}
For the thermodynamics sufficiently large number $N$ of particles, it is sufficient to evaluate the density of states in the semiclassical approximation. Then, the energy surface is defined by
\bey
E = \frac{p^2}{2m} + \frac{{\bf k}^2}{2m} + mgx,
\eey
where $p$ is the momentum in the $x$ direction, and ${\bf k}$ is the momentum in the directions normal to $k$.

The number-of-states function is
\bey
\Omega(E) = \frac{A}{4\pi^3} \int d^2 k   \int dx  \sqrt{2mE - k^2 -2m^2gx}.
\eey
Carrying out the integration with respect to $k$, we obtain
\bey
\Omega(E) &=& \frac{2^{3/2}A}{3\pi^2} m^3 g^{3/2} \times \left\{ \begin{array}{cc} \int_{\ell_b}^{\ell_t} dx \left[\frac{E}{mg} - x\right]^{3/2}, & E > mg\ell_t \nonumber 
\\
\int_{\ell_b}^{\frac{E}{mg}}dx \left[\frac{E}{mg} - x\right]^{3/2} &
mg\ell_b \leq E \leq mg\ell_t \end{array} \right. \\
&=& \frac{4\sqrt{2m}A}{15\pi^2 g} \times \left\{ \begin{array}{cc} (E - mg\ell_b)^{5/2} - (E - mg\ell_t)^{5/2}, &  E > mg\ell_t 
\\ (E - mg\ell_b)^{5/2},& mg\ell_b \leq E \leq mg\ell_t. \end{array} \right.
\eey
Eq. (\ref{rhoe}) follows from differentiation of $\Omega(E)$ with respect to the energy $E$.

\section{Thermodynamic properties of massless relativistic gases in Rindler spacetime}
 \renewcommand{\theequation}{B-\arabic{equation}}
  \setcounter{equation}{0}

\subsection{Classical gas}
From Eq. (\ref{pfz0}), for  $m = 0$, we calculate the internal energy $U = - \partial \log Z/\partial \beta$ and the gravitational pull $Q = -  \partial \log Z/\partial \eta$. We choose coordinates so that $\ell_b = 0$, in order to make sure that no value of $L$ crosses the Rindler horizon. Setting 
 $\eta = g \beta$, we find
\bey
U = \frac{3N}{\beta} \frac{1 +  g L + \frac{1}{3} g^2L^2}{(1+gL)(1 + \frac{1}{2} gL)},\\
Q = \frac{N}{g\beta }\left(1 - \frac{1}{ (1+gL)(1 + \frac{1}{2} gL)}\right).
\eey
Next, we calculate the pressures. Since we fixed $\ell_b = 0$, variation with respect to $L$ yields the top pressure $P_t$. To compute the bottom pressure $P_b$, we add $Ug/a$ to $P_t$. We obtain
\bey
P_h = \frac{N}{AL\beta}, \hspace{1cm} P_t =  \frac{P_h}{ (1+gL)(1 + \frac{1}{2} gL)}, \hspace{1cm} P_b = P_h \frac{1 +\frac{9}{2}gL + \frac{7}{2} g^2L^2 +g^3L^3}{(1+gL)(1 + \frac{1}{2} gL)}.
\eey
The three pressures satisfy $P_t \leq P_h \leq P_b$. Equality is achieved for $gL = 0$. At $gL \rightarrow \infty$,  $P_t$ and $P_h$ vanish
and   $P_b$ diverges.

\subsection{Quantum electromagnetic field}
From Eq. (\ref{photon}), we calculate the internal energy $U$ and the gravitational pull for a gas of photons
\bey
U &=& \frac{4 \pi^2 AL}{15 \beta^4} \frac{1 +gL + \frac{1}{3} g^2L^2}{(1 + gL)^3}. \\
Q &=& \frac{2\pi^2 AL^2}{15\beta^4} \frac{1 + \frac{1}{3}gL}{(1 + gL)^3} = \frac{UL}{2} \frac{1 + \frac{1}{3}gL}{1 +gL + \frac{1}{3} g^2L^2}.
\eey
We note that for $gL \rightarrow 0$, the internal energy given by Planck's law, and $Q = {L}/{2}$. For $gL \rightarrow \infty$, $U$ is $L$-independent, and it scales with the area $A$. In this regime, $Q = {U}/({2g})$; the gravitational pull also scales with area. The gravitational susceptibility is
\bey
\chi_T = \frac{32 \pi^2 AL^3}{45 \beta^4}\frac{1  + \frac{1}{4} gL}{(1 + gL)^4}
\eey

We also evaluate the pressures:
\bey
P_h &=& \frac{4 \pi^2 }{45 \beta^4} \frac{1+\frac{1}{2}gL}{(1+gL)^2} = \frac{1}{3} \frac{U}{AL} \frac{(1+gL)(1+\frac{1}{2}gL)}{1 +gL + \frac{1}{3} g^2L^2}\\
P_t &=& \frac{4\pi^2}{45\beta^4} \frac{1}{(1+gL)^3} = \frac{1}{3} \frac{U}{AL} \frac{1}{1 +gL + \frac{1}{3} g^2L^2} \\
P_b &=& \frac{4\pi^2}{45\beta^4} (1+gL) = \frac{1}{3} \frac{U}{AL}  \frac{(1+gL)^4}{1 +gL + \frac{1}{3} g^2L^2}
\eey
The three pressures satisfy $P_t \leq P_h \leq P_b$. They all
equal  $\resizebox{0.8cm}{0.45cm}{$\dfrac{1}{3}\dfrac{U}{AL}$}$ as $gL \rightarrow 0$. However, $P_t$ and $P_h$ vanish as $gL \rightarrow \infty$, while   $P_b$ diverges.

\end{appendix}

\begin{thebibliography}{}



\bibitem{Bek1}  J. D. Bekenstein, {\em Black holes and entropy}, Phys. Rev. D  {\bf 7}, 8 (1973). 


\bibitem{Bek2}  J. D. Bekenstein, {\em Generalized second law of thermodynamics in black hole physics}, Phys. Rev. D  {\bf 9}, 3292  (1974). 
 

\bibitem{BCH}  J. M. Bardeen, B. Carter, and S. W. Hawking, {\em The four laws of black hole mechanics}, Commun. Math. Phys. {\bf 31}, 161  (1973). 


\bibitem{Hawk1}  S. W. Hawking, {\em Particle creation by black holes}, Comm. Math. Phys.  {\bf 43}, 199  (1975). 

\bibitem{Wald} R. M. Wald, {\em The Thermodynamics of Black Holes}, Living Rev. in Rel. 4, 6 (2001). 

\bibitem{Penrose} R. Penrose, {\em Singularities and time-asymmetry}, in ``Einstein Centenary Volume", S. W. Hawking and W. Israel (eds.), (Cambridge: Cambridge
University Press, 1979).

\bibitem{Wald2} R. M. Wald, {\em The Arrow of Time and the Initial Conditions of the Universe}, Stud. Hist. Philos. Mod. Phys.   37, 394 (2006). 

\bibitem{Wallace} D. Wallace, {\em  Gravity, entropy, and cosmology: In search of clarity},  Br. J. Philosophy Sci.   61, 513 (2010).

\bibitem{Kron}Ø. Grøn,  {\em Entropy and Gravity}, Entropy  14, 2456 (2012).



\bibitem{Pad} T. Padmanabhan, {\em Statistical mechanics of gravitating systems}, Phys. Rep. 188, 285 (1990).

\bibitem{Katz} J. Katz, {\em Thermodynamics of self-gravitating systems}, Found. Phys. 33, 223 (2003).

\bibitem{Chavanis} P.H. Chavanis, C. Rosier, and C. Sire, {\em Thermodynamics of self-gravitating systems }, Phys. Rev.  E66, 036105 (2002).

\bibitem{AnSav14} C. Anastopoulos and K. Savvidou, {\em The thermodynamics of self-gravitating systems in equilibrium is holographic}, Class. Quant. Grav. 31,  055003 (2014).

\bibitem{Caratheo} C. Carathéodory,  {\em Untersuchungen über die Grundlagen der Thermodynamik},  Mathematische Annalen. 67,   355 (1909).

\bibitem{Landsberg} P.T. Landsberg, {\em Foundations of thermodynamics}, Rev. Mod. Phys. 28, 363 (1956).

\bibitem{Giles} R. Giles, {\em Mathematical Foundations of Thermodynamics} (New York: Pergamon Press, 1964).



\bibitem{Callen} H. B. Callen,  {\em Thermodynamics and an Introduction to Thermostatistics} (John Wiley, New York, 1985).


\bibitem{Lieb}  E. H. Lieb and J. Yngvason, {\em The physics and mathematics of the second law of thermodynamics}, Phys. Rep.  {\bf 310}, 1 (1999). 



\bibitem{SouAn23}  E. Sourtzinou and C. Anastopoulos, {\em Quantum statistical mechanics near a black hole horizon}, Phys. Rev. D107, 085006 (2023). 

\bibitem{Souphd} E. Sourtzinou, {\em Thermodynamics in a gravitational field}, Ph.D. dissertation (University of Patras). URI: \url{https://hdl.handle.net/10889/28224}
  
\bibitem{UW} W. G. Unruh and R. M. Wald, {\em Acceleration radiation and the generalized second law of thermodynamics}, Phys. Rev. D25, 4 (1982). 

 \bibitem{Martin1} E. A. Martinez, {\em The postulates of gravitational thermodynamics}, Phys. Rev. D54, 6302 (1996).
 
\bibitem{Martin2} E. A. Martinez, {\em Fundamental thermodynamical equation of a self-gravitating system},  	Phys. Rev. D53,  7062  (1996).

  \bibitem{AnSav12} {C. Anastopoulos and N. Savvidou, {\em Entropy of singularities in self-gravitating radiation}, Class.
Quant. Grav.   29 , 025004 (2012).}


\bibitem{Kotop1}  D. Kotopoulis and C. Anastopoulos, {\em Thermodynamics and phase transitions of black holes in contact with a gravitating heat bath}, Class. Quant. Grav.  38 ,  195026 (2021). 




\bibitem{LDP} P. T. Landsberg, J. Dunning Davies, and D. Pollard, {\em Entropy of a column of gas under gravity}, Am. J. Phys.   62 , 712 (1994). 




\bibitem{BiTr} J. Binney and S. Tremaine, {\em Galactic Dynamics} (Princeton Series in Astrophysics, 1987).


\bibitem{Chandr} S. Chandrasekhar, {\em An Introduction to the Theory of Stellar Structure} (Dover, 1942).



    


\bibitem{Callen2} H. B. Callen, {\em Thermodynamics as a science of symmetry}, Found Phys 4, 423 (1974).


\bibitem{GrMa} S.R. de Groot and P. Mazur, {\em Non-equilibrium thermodynamics} (New York: Dover, 1982).

\bibitem{CoLa} C. A. Coombes and H. Laue, {\em A paradox concerning the temperature distribution of a gas in a gravitational field}, Am. J. Phys. 53, 272 (1985).

\bibitem{Gibbs} J. W. Gibbs, {\em On the equilibrium of heterogeneous substances} in ``The Scientific Papers of J. Willard Gibbs - Volume One" (Oxford: Oxbow Press, 1993).


\bibitem{RWV} F. L. Román,   J. A. White, and S. Velasco, { \em Microcanonical single-particle
distributions for an ideal gas in a
gravitational field}, Eur. J. Phys. 16, 83  (1995). 


\bibitem{RGWV} F. L. Román, A. González,   J. A. White, and S. Velasco, { \em Microcanonical ensemble study of a gas column under gravity}, Z. Phys. B 104, 353 (1997). 

\bibitem{Jaynes} E. T. Jaynes, {\em Probability Theory: The Logic of Science} (Cambridge: Cambridge University Press, 2003). 

\bibitem{Castel} G. Castellano, {\em Thermodynamic potentials for simple magnetic systems}, J. Magn. Magn. Mater. 260, 146 (2003).

\bibitem{SaBa} S. Samanta and  B. R. Majhi,
{\em A note on entropy of matter in presence of gravity: Status of extensivity of entropy}, Ann. Phys.
  481,
170161 (2025).

\bibitem{Salz} A. M. Salzberg, {\em Exact Statistical Thermodynamics of Gravitational Interactions in One and Two Dimensions}, J. Math. Phys. 6, 158 (1965).

\bibitem{Rybicki}G. Rybicki,{\em  Exact Statistical Mechanics of a One-Dimensional Self-Gravitating System}, Astrophys. Space. Sci 14, 56 (1971).

\bibitem{Mann1}T. Ohta and R.B. Mann, {\em Canonical reduction of two-dimensional gravity for particle dynamics}, Class. Quant. Grav. 13, 2585 (1985).

\bibitem{Mann2} R. B. Mann and P. Chak, {\em Statistical mechanics of relativistic one-dimensional self-gravitating systems}, Phys. Rev. E65, 026128 (2002).


\bibitem{LL}L. D. Landau and E. M. Lifshitz, {\em Statistical physics}  (Oxford: Pergamon Press, 1980).

\bibitem{Huang} K. Huang, {\em Statistical mechanics} (New York: Wiley, 1991).


\bibitem{LMar}  D. J. Louis-Martinez, {\em Classical relativistic ideal gas in thermodynamic equilibrium in a uniformly accelerated reference frame}, Class. Quantum Grav. 28, 035004 (2011). 
    
    \bibitem{Pad2}  S. Kolekar and T. Padmanabhan, {\em Ideal gas in a strong gravitational field: Area dependence of entropy}, Phys. Rev. D83, 064034 (2011). 
    
    \bibitem{Pad3} S. Bhattacharya, S. Chakraborty, and T. Padmanabhan, {\em Entropy of a box of gas in an external gravitational field revisited }, Phys. Rev. D96, 084030  (2017).

    \bibitem{Kole} I. Kulikov, {\em Thermodynamics of trapped fermions in gravitational field}, Phys. Lett. A286,  25 (2001).

    \bibitem{Thai}  W. Kanchanapusakit  and P. Tanalikhit, {\em An Ideal Fermi Gas Under Uniform Gravity}, SSRN preprint, http://dx.doi.org/10.2139/ssrn.5279665.

\bibitem{BBB} J. D. Bekenstein, {\em Universal upper bound on the entropy-to-energy ration for bounded systems}, Phys. Rev. D23, 287 (1981).

\bibitem{BGSL} J. D. Bekenstein, {\em Entropy bounds and the second law for black holes}, Phys. Rev. D27, 10, (1983).

\bibitem{barometric} M. N. Berberan-Santosa, E. N. Bodunov, and  L.Pogliani, {\em On the barometric formula}, J. Phys. 65, 404 (1997).

\end{thebibliography}
\end{document}